# Direct observations of a complex coronal web driving highly structured slow solar wind

L. P. Chitta[1]*, D. B. Seaton[2], C. Downs[3], C. E. DeForest[2] & A. K. Higginson[4]

[1]Max Planck Institute for Solar System Research, Justus-von-Liebig-Weg 3, 37077 Göttingen, Germany
[2]Southwest Research Institute, 1050 Walnut Street, Suite 300, Boulder, CO 80302, USA
[3]Predictive Science Inc., 9990 Mesa Rim Road, Suite 170, San Diego, CA 92121, USA
[4]NASA Goddard Space Flight Center, 8800 Greenbelt Rd, Greenbelt, MD 20771, USA

*To whom correspondence should be addressed; E-mail: chitta@mps.mpg.de.

**The solar wind consists of continuous streams of charged particles that escape into the heliosphere from the Sun, and is split into fast and slow components, with the fast wind emerging from the interiors of coronal holes. Near the ecliptic plane, the fast wind from low-latitude coronal holes is interspersed with a highly structured slow solar wind, the source regions and drivers of which are poorly understood. Here we report extreme-ultraviolet observations that reveal a spatially complex web of magnetized plasma structures that persistently interact and reconnect in the middle corona. Coronagraphic white-light images show concurrent emergence of slow wind streams over these coronal web structures. With advanced global MHD coronal models, we demonstrate that the observed coronal web is a direct imprint of the magnetic separatrix web (S-web). By revealing a highly dynamic portion of the S-web, our observations open a window into important middle coronal processes that appear to play a key role in driving the structured slow solar wind.**

Historically, the solar wind[1] has been categorized as "fast" or "slow" wind. The fast wind is generally described as having speeds greater than 500 km s$^{-1}$ and originating from interiors of coronal holes formed by open magnetic flux. The slow wind is classified as wind with speeds less than 500 km s$^{-1}$ and is associated with the coronal streamer belt[2]. In general, slow solar wind is characterized its high degree of internal structure and variability, coronal elemental compositions, and ionic compositions that indicate a hotter and denser source region than the fast wind[3–7]. Interplanetary observations reveal two components of slow wind, one that is characterized by its high degree of Alfvénic fluctuations similar to the fast wind, and the other with lower Alfvénicity[8].

During solar minimum, the coronal streamer belt and its associated slow wind streams remain concentrated around the solar equatorial regions, but during periods of solar activity, small coronal holes, which quite commonly form in the unipolar remnants of active regions confined to low latitudes, can reshape the streamer belt, extending it to high latitudes and distorting the central heliospheric current sheet (HCS). These complex streamer belt topologies must create a web of magnetic separatrices within the corona, termed the S-web[9], but how exactly the coronal streamer belt and this coronal S-web relate to the origin of the slow solar wind is still a subject of active debate.

Super-radial expansion of the coronal holes themselves is thought to be an important source of slow solar wind[10–15]. MHD wave turbulence could drive the slow wind along open magnetic fields in coronal holes[16]. In this scenario, variations in expansion factors and footpoint field strengths in open field regions are thought to play a role, at least in part, in the observed compositional difference between fast and slow winds[11]. The spatial variability of slow solar wind is attributed to large-scale events such as coronal mass ejections and streamer blobs that propagate through the background slow solar wind[11].

At the same time, interchange reconnection between open and closed magnetic fields[17], and S-web reconnection (i.e. at quasi-separatrix layers and separatrices) created by low-latitude coronal holes[18] are also commonly invoked drivers of the slow solar wind and/or its internal structure. In this scenario, at least a component of the structured slow wind is attributed to be driven by reconnection that is inherently dynamic and transient. Such models also predict that S-web dynamics are particularly important within the middle corona, a transitional regime where magnetic topology changes from being predominantly closed to open, with altitude. In streamer belt regions, this transition typically occurs over heights of 1.5 solar radii (from the Sun's center; $1R_\odot$ ~ 700 Mm. Magnetic reconnection in the middle corona could release plasma from the closed magnetic structures into the solar wind[5,18]. But the middle coronal manifestation of reconnection dynamics associated with coronal holes and active regions, and the extent to which they drive slow solar wind structures, remain poorly understood[19].

## Results

We used images from a special off-pointing campaign of GOES Solar Ultraviolet Imager (SUVI)[20,21], that directly observed the middle corona, to identify coronal dynamics associated with the highly structured slow solar wind. The extended field of view of SUVI allowed us to trace extreme ultraviolet (EUV) features up to a height of about $2.7R_\odot$, into the middle corona (Methods). Using its EUV 195 Å passband, SUVI observed a pair of roughly east-west aligned coronal holes near the equator with an embedded (decaying) active region. During the course of one month of observations, we observed this coronal hole-active region (CH-AR) system as it rotated across the solar disk and subsequently when it appeared at the west and east limbs. The limb view of the system revealed a highly structured middle corona with elongated features, which we term a "coronal web", that protrudes radially outward in the plane of sky, with a large latitudinal extent (Fig. 1 and Extended Data Fig. 1).

This coronal web exhibited complex dynamics. In Fig. 1a-b, we show a SUVI snapshot up to a height of about $2.7R_\odot$ (bounded by solid white arc), from the period when the CH-AR system was at the west limb. Above this height, the SUVI EUV image is paired with a co-temporal white light coronagraphic image from the Large Angle Spectroscopic Coronagraph (LASCO)[22] onboard the SOHO spacecraft (Methods). We point to an outflowing stream on the west-limb above $2.65R_\odot$, observed with SOHO/LASCO (arrow in panels a and b). Examining the image sequence reveals that similar streams are repeatedly driven in that region (Fig. 1c highlights two such events). Underlying these streams, in the middle corona, we found a system of interacting magnetic structures that are clearly linked to the CH-AR system, in SUVI images. Magnetic structures that are apparently open, initially, approach each other and interact to form (transient) closed loops; two such interactions associated with stream events in Fig. 1c are displayed in Fig. 1d (see Supplementary Video 1). Three other examples of solar wind streams emerging over interacting middle coronal structures are presented in Supplementary Figures 1–3 (Supplementary Videos 2–4). We found that the speeds of these outflows are consistent with speeds of slow solar wind streams (Supplementary Figure 4 and Supplementary Information). Such streams are generally considered to be passive tracers of the slow solar wind[23,24]. We also found that these types of events are quite prevalent in the coronal web. Over the course of five days when the system crossed the west limb, we observed persistent interaction and continual rearrangement of the coronal web features, and emergence of highly structured and variable slow solar wind streams over their tips (Supplementary Video 5). We observed similar (plane of sky) complex dynamics in the middle coronal web and the emergence of slow solar wind streams also when the same system crossed the east limb (Supplementary Figure 5; Supplementary Video 6).

Based on low-coronal EUV images and photospheric magnetic field data from the Solar Dynamics Observatory[25], we found that the coronal web is associated with a highly warped heliospheric current sheet (HCS; Methods, Extended Data Figs. 2-3; Supplementary Information). To decipher the complexity of the latitudinally extended complex coronal web and to investigate how the CH-AR system is topologically related to the global magnetic field, we employ advanced global 3D MHD coronal models driven at the lower boundary by synoptic observations of photospheric magnetic fields[26,27] (Methods). We visualize the skeleton of the global magnetic field using the squashing factor, $Q$, which employs field line-mappings to identify the quasi- and true-separatrix surfaces where $Q \gg 1$ (ref.[28]). In particular, the synoptic map of the signed $\log Q$ at $3R_\odot$, colored blue(red) by the inward(outward) direction of magnetic field, illustrates complex S-web features[29], that map from the coronal holes at the surface to the open middle-corona and warped HCS above the active region (Fig. 2a and Extended Data Figs. 4-5). This signed $\log Q$ map shows longitudinally extended, prominent complex S-web features surrounding the active region. These include a warped global polarity inversion line along with a pair of northeast and southwest pseudostreamer arcs on larger scales, with embedded cellular pockets of smaller scale features which are formed by separatrices in the mapping due to presence of distinct small-scale flux concentrations at the surface within the larger open field region (Supplementary Information).

As the CH-AR system crossed the west-limb, we first observed the signature of the polarity inversion line in the form of a bipolar helmet streamer, northwest of the active region. Similarly, in the southwest, we observed a unipolar pseudostreamer (Fig. 2b; symbols on line-b in Fig. 2a). Then we observed this configuration gradually changing to a unipolar pseudostreamer in the northeast and a bipolar helmet streamer in the southeast with solar rotation (Fig. 2c; symbols on line-c in Fig. 2a). More importantly, through this entire period (about five days), SUVI recorded complex coronal web activity over the CH-AR system, along with observations of solar wind streams in LASCO (Supplementary Video 5). Similar middle coronal dynamics are observed at the east-limb (Supplementary Figure 5; Supplementary Video 6). Its observed persistence suggest that the coronal web is not only extended latitudinally (Extended Data Fig. 1), but also spans longitudinally over the entire CH-AR system on either side of the HCS (Fig. 2d).

Thus, our SUVI observations captured direct imprints and dynamics of this S-web, in the middle corona. For instance, consider wind streams presented in Fig. 1. Those outflows emerge when a pair of middle coronal structures approach each other. By comparing the timing of these outflows in Supplementary Video 5, we found that the middle coronal structures interact at the cusp of the southwest pseudostreamer. Similarly, wind streams in Supplementary Figures 1–3 emerge from the cusps of the HCS. Models suggest that streamer and pseudostreamer cusps are sites of persistent reconnection[30,31]. The observed interaction and continual rearrangement of the coronal web features at these cusps are consistent with persistent reconnection, as predicted by S-web models. Although reconnection at streamer cusps in the middle corona has been inferred in other observations studies[32,33], and modeled in 3D[30,31], the observations presented here represent imaging signatures of coronal web dynamics and their direct and persistent effects. Our observations suggest that the coronal web is a direct manifestation of the full breadth of S-web in the middle corona. The S-web reconnection dynamics modulate and drive the structure of slow solar wind through prevalent reconnection[9,18].

A volume render of $\log Q$ highlights the boundaries of individual flux-domains projected into the image plane, revealing the existence of substantial magnetic complexity within the CH-AR system (Fig. 3a; Supplementary Video 7). The ecliptic view of the 3D volume render of $\log Q$ with the CH-AR system at the west-limb does closely reproduce elongated magnetic topological structures associated with the observed coronal web, confined to northern and southern bright (pseudo-) streamers (Fig. 3b; Supplementary Video 8). The synthetic EUV emission from the inner to middle corona, and the white-light emission in the extended corona (Fig. 3c) are in general agreement with structures that we observed with SUVI-LASCO combination (Fig. 1a). Moreover, radial velocity sliced at $3R_\odot$ over the large-scale HCS crossing and the pseudostreamer arcs in the MHD model, also quantitatively agrees with the observed speeds of wind streams emerging from those topological features (Supplementary Figures 4 and 6 and Supplementary Information). Thus, the observationally-driven MHD model provides credence to our interpretation of the existence of the complex coronal web whose dynamics correlate to the release of wind streams.

The long lifetime of the system allowed us to probe the region from a different view point using sun-orbiting STEREO-A, that was roughly in quadrature with respect to the Sun-Earth line during SUVI campaign (Methods; Extended Data Fig. 6). By combining data from STEREO-A's extreme ultraviolet imager (EUVI)[34], outer visible-light coronagraph (COR-2), and the inner visible-light heliospheric imager (HI-1)[35], we found imprints of the complex coronal web over the CH-AR system extending into the heliosphere. Fig. 4a and the associated Supplementary Video 9 demonstrate the close resemblance between highly structured slow solar wind streams escaping into the heliosphere and the S-web-driven wind streams that we observed with SUVI and LASCO combination. Due to the lack of an extended field of view, the EUVI did not directly image the coronal web that we observed with SUVI, demonstrating that the SUVI extended field-of-view observations provide a crucial missing link between middle coronal S-web dynamics and the highly structured slow solar wind observations.

## Discussion

Our observations provide important close-to-the-sun context to the measurements of the highly structured nature of slow solar wind emerging from an equatorial coronal hole recorded by Parker Solar Probe (PSP) during its first perihelion on 2018-11-06 (ref.[7]). The western coronal hole that we discussed here is the same region to which PSP was magnetically connected to during its perihelion at a later date[7]. A similar north-south aligned polarity inversion line separated the western and eastern coronal holes and the active region which was decaying at that time (Methods; Extended Data Figs. 3 and 6). Some 10 days before PSP's encounter, instruments on STEREO-A recorded inner coronal and heliospheric structures similar to those found during SUVI observations (Figs. 4b; Supplementary Video 10). During its perihelion, PSP observed ubiquitous radial velocity spikes in the solar wind, associated with switchbacks, local reversals of radial magnetic fields[7]. Such switchbacks are linked to the inherently Alfvénic slow wind originating from rapidly expanding small coronal holes[7,15]. Switchbacks could also form in situ in the expanding solar wind[36]. At the same time, models predict that interchange reconnection between open and closed magnetic fields, such as in the S-web, could generate and launch Alfvén waves[30] and magnetic flux ropes[37], which in turn may further evolve to form switchbacks.

Both photospheric and low coronal observations point to ubiquitous small-scale ephemeral magnetic activity in and around active regions and in coronal holes[38,39]. Such small-scale magnetic activity will lead to separatrices even in open field coronal holes near active regions. Our MHD model does reveal more cellular and smaller separatrices on either side of the active region (Fig. 2a), which result from small-scale magnetic concentrations within the observed coronal holes (Methods). Although origin of switchbacks still remains elusive, our SUVI observations and MHD model hint that the S-web interchange reconnection could play a role in the process. Recent observations do hint at such a close link between interchange reconnection and switchbacks[40,41].

Systems of low-latitude coronal holes surrounding active regions are quite common during periods of solar activity[13]. The CH-AR system presented here is well-isolated and thus allowed us to observed full breadth of S-web dynamics in the middle corona. Because such systems are common, based on our SUVI observations, we suggest that a complex coronal web is quite common, too.

Our findings lend support to reconnection-based slow solar wind models[9,17,18], and identify two critical components necessary for further study of the coronal S-web. Firstly, depending on the alignment of remote observatories and the orientation of global polarity inversion line, the S-web imprints that we clearly observed with SUVI may not be apparent in the plane of sky due to projection effects, but that does not necessarily imply it is absent. Moreover, single vantage point middle coronal observations do not provide a detailed 3D view of the reconnection. In addition, uncertainties that arise due to superposition of structures in the middle corona cannot be resolved in such single vantage point observations. Here, a polar mission targeting the middle corona in combination with ecliptic view could, one day, reveal the 3D structure of the S-web. Secondly, as demonstrated by Figs. 1 and 4, above the closed field regions the S-web separatrices are more established and less likely to change due to topology. A direct consequence of this can be seen in the predominantly radial flows that are ubiquitous beyond 3–4$R_\odot$. Therefore, without direct images of the middle corona important and more complex S-web dynamics associated with the solar wind, which are evidently more frequent in this regime as compared to the outer corona, will continue to go undetected. Additionally, because the middle corona provides direct access to the global topological skeleton of Sun's magnetic fields (Fig. 2), observing this regime provides crucial information on how plasma upflows that originate low in the corona eventually make their way through the relatively complex topology in the middle corona and into the solar wind.

Using the long-duration EUV image sequence of the solar corona up to 2.7$R_\odot$, complemented by observationally-driven advanced 3D MHD coronal models, we identified a complex coronal web as a direct imprint of the S-web. While the speed of bulk of the slow solar wind itself could be closely governed by the expansion factors of coronal holes[11–13], our observations suggest that S-web dynamics imprint magnetic topological structures on the slow wind and play a role in releasing hotter plasma with coronal compositions into the slow wind[5]. The dynamic evolution of the observed coronal web thus presents direct observational evidence for a key and integral role of S-web in driving of the highly structured slow solar wind.

Our SUVI observations are limited in both time and spatial resolution that would be needed to fully resolve the origins of slow solar wind and switchbacks. To this end, our exploratory measurements of the time-dependent middle corona lay a clear path forward. They demonstrate the need for a more systematic and detailed EUV observations of the middle coronal region in close coordination with the outer coronal and in situ measurements. Future and planned missions, including NASA's in-development PUNCH mission[42] will partially address this need, for instruments operating from the Earth's perspective. Additional missions proposed and in development that will explore the middle corona specifically, such as ECCCO[43], SunCET[44], PROBA3/ASPIICS[45] among others will help to fully connect the global corona-heliosphere system. Combined with 3D information we can derive from PSP[46], Solar Orbiter's Metis[47] and Extreme-Ultraviolet Imager[48] away from the Sun-Earth line, and new spectral capabilities in several proposed missions, we can address these concerns to fully close these questions.

## Methods

## Observations

We used observations from four different spacecrafts to gain a broader view of coronal sources and to trace origins of the highly structured slow solar wind. These observations span a range of distances from the Sun, covering features at low altitudes closer to the solar surface to distances exceeding 20 times the solar radius (1$R_\odot$ ~ 700 Mm).

**GOES SUVI Data.** The GOES Solar Ultraviolet Imager (SUVI) data we used come from a deep-exposure offpoint campaign run between 2018-08-07 and 2018-09-13 (ref.[20]) with two extreme ultraviolet (EUV) filters: 171 Å and 195 Å. The SUVI observations are three-panel mosaic images, with a central sun-centered image and two eastward and westward off-pointed side panels[49]. SUVI was not designed to be operated in off-pointed mode, so some additional processing is required to reduce image noise, remove background stray light, and reduce the large-scale brightness gradient (up to a factor of $10^6$ between the brightest and faintest structures) to permit the visualization of the whole field of view in a single, uniformly scaled image. The resulting images have a field of view out to about $5R_\odot$, at an image scale of 5″ pixel$^{-1}$, and a cadence of around 20 minutes. Because of the significant decrease in EUV emission at large heights, and the significant contribution of noise resulting from stray light, there is negligible structure detectable above $3.5$–$4R_\odot$ in these observations. Additional details about the processing, contents, and limitations of these observations appear in ref.[20].

In Extended Data Fig. 1 we present the overview of SUVI coronal observations recorded by its 195 Å filter. The period of SUVI observations covers a part of the decay phase of previous solar activity cycle (number 24). During the course of approximately one month of observations, SUVI observed multiple coronal holes and active regions. Among these features are a pair of coronal holes, roughly aligned along the east-west orientation. The eastern coronal hole appears to be an extension of the northern polar coronal hole and the western coronal hole is more isolated. A decaying active region (NOAA number 12711) is embedded in this pair of coronal holes. Based on the observations from the Solar Dynamics Observatory we identified that these two coronal holes and the active region lived for at least five solar rotations (i.e., about 135 days). In the lower panels of Extended Data Fig. 1 we mark these features when they appear on the west-limb (panel b) and then on the east-limb after rotating through the far-side of the Sun (panel c).

**SOHO LASCO Data.** To extend the field of view covered by SUVI to larger heights, and fill in gaps where the EUV emission revealed by it is insufficient to detect structure, we use data from the Large Angle Spectroscopic Coronagraph (LASCO)[22] C2 coronagraph, which has a field of view that extends roughly to $6R_\odot$. We prepared these data using the standard LASCO data reduction tools distributed in the Interactive Data Language (IDL) SolarSoft by the LASCO team. We improve signal-to-noise in these images by stacking in time to reach an effective cadence of one hour. We used a custom calculated minimum background spanning the entire observation set to remove stray light and F-corona, and an additional despiking step to suppress both cosmic rays and background stars. Due to the inherent significant differences in the intrinsic brightness of the visible light corona observed by LASCO and the EUV emission observed by SUVI, to merge these LASCO observations with SUVI we also must rescale the data to approximately the same normalized dynamic range. We do this by applying a radial normalizing filter, derived from the median of the whole data over time. We then merge the data sets by pairing images on the basis of proximity in time. The resulting SUVI-LASCO composites show EUV emission up to about $2.7R_\odot$ and the visible light corona above this. Additional details on the processing and merger of these data sets appear in ref.[20].

**SDO Data.** Solar Dynamics Observatory (SDO)[25] provided lower-coronal and photospheric context of the coronal hole-active region (CH-AR) system. We have used the 171 Å, 193 Å, and 211 Å EUV filter full-disk images from the Atmospheric Imaging Assembly (AIA)[50] onboard SDO to examine the general evolution of coronal structures over the course of one month from 2018-08-08 to 2018-09-09 (roughly covering the period of SUVI off-pointing campaign). The surface magnetic structures were observed using the full-disk line-of-sight magnetic field maps obtained by the Helioseismic and Magnetic Imager (HMI)[51] on SDO. We retrieved these full-disk AIA and HMI data at a 2-hour cadence from the Joint Science Operations Center (JSOC), which are then processed using the standard AIA_PREP procedure available in SolarSoft.

We visually identified the system of eastern and western coronal hole pair and the embedded active region on 2018-08-10 and followed those points at the rate of Carrington rotation to demonstrate that it is indeed the same set of features that rotate to the west limb around 2018-08-16 and then reappear on the east limb after rotating through the far-side of the Sun around 2018-08-31 (Extended Data Fig. 2). Closer to the eastern and western limbs, the line-of-sight component of surface magnetic field patches associated with the CH-AR system may not be apparent in the HMI maps. Here, the system is better seen in the coronal images in which the associated active region loops and coronal holes are evident. The active region embedded between the two coronal holes emerged on 2018-05-23 and received the NOAA number 12711 the next day. The active region is composed of leading negative polarity magnetic fields and trailing positive polarity magnetic fields. The western coronal hole formed within a region of predominantly negative polarity magnetic field, whereas the eastern coronal formed in a region of predominantly positive polarity magnetic field.

**AIA synoptic maps.** We complemented the full-disk SDO images with the AIA 171 Å, 193 Å, and 211 Å EUV filter synoptic Carrington maps[52] to investigate the general long-term evolution of the pair of coronal holes and the active region. Each synoptic map covers a full Carrington rotation (CR) and are in Carrington-longitude and Sine-latitude projection. In particular, we used AIA synoptic maps from CR2206 through CR2211 (from 2018-07-09 to 2018-12-20). These CRs cover the period of the SUVI 2018 off-pointing campaign (2018-08-07 to 2018-09-13), as well as the first perihelion of Parker Solar Probe on 2018-11-06 (ref.[7]) (see Extended Data Fig. 3).

The coronal hole pair exhibited evolution throughout the period spanning the six Carrington rotations that we considered, both in terms of their spatial morphology and general location with respect to their position that we visually determined on 2018-08-10. But the pair remained clearly distinguishable. The active region, on the other hand, exhibited a continual decay from CR2206 through CR2210 and was nearly absent or completely decayed in CR2211. To illustrate these long-term changes in the lower-coronal features, we visually identified and marked their new positions in each CR. All the features migrated westward (i.e., increasing Carrington longitudes). For example, the Carrington longitude of the western coronal hole in CR2206 was roughly 320°, whereas in CR2211, it was 350°.

**HMI synoptic maps for magnetic field extrapolations.** The magnetic polarity inversion line (PIL) will provide information on the general distribution and separation of domains of positive and negative polarity magnetic fields at a given height above the solar surface. This provides further insights into the observed (large-scale) coronal plasma structures that generally outline magnetic fields. To this end, we employed a potential field source surface (PFSS) technique to extrapolate coronal magnetic fields from the observed surface magnetic field maps[53]. As the lower boundary to the PFSS model, we used synoptic surface radial magnetic field maps archived at JSOC (CRs2206–2211). Specifically, we used synoptic maps of the type hmi.synoptic_mr_polfil_720s, in which the polar fields above 60° latitudes are inferred from annual observations of the poles when the respective north or south pole of the Sun is most tipped toward Earth[54]. More details about this data product are available at http://hmi.stanford.edu/QMap/. For these magnetic field extrapolations, we used a PFSS code that is distributed in the SolarSoft. We placed the source surface or the top boundary at a height of $2R_\odot$ (ref.[7]). Above this height or source surface, magnetic fields are assumed to remain open. The contour where the radial magnetic field is zero at the source surface defines the PIL, which is shown as the yellow line in Extended Data Fig. 3.

**STEREO Data.** During 2018 August, STEREO-A spacecraft[55] was orbiting the Sun roughly in quadrature with respect to the Sun-Earth line at Earth Ecliptic longitude of about -108° (in the Heliocentric Earth ecliptic coordinate system). This rough quadrature means that the solar features in Earth's line of sight would appear on the western limb of the Sun as seen from the vantage point of STEREO-A. The CH-AR system traversed the solar disk and the central meridian (from Earth's perspective) between 2018-08-07 and 2018-08-12, the period that is roughly one week before when it rotated to the western limb as observed with both SUVI and SDO (see Extended Data Figs. 1-2). This enabled us to probe the heliospheric manifestation of the CH-AR system when it was approximately in Earth's line of sight, using images obtained by a suite of instruments on the Sun Earth Connection Coronal and Heliospheric Investigation (SECCHI) observatory[34] on STEREO-A.

In particular, we employed coronal images obtained with the 195 Å filter on the extreme ultraviolet imager on SECCHI (EUVI; that records EUV emission from coronal plasma around 1–2 MK in the inner corona up to a field of view of about $1.7R_\odot$). These images are combined with imaging data from SECCHI's outer Lyot visible-light coronagraph (COR2; covering the extended corona between 2.5 to $15R_\odot$) and the inner visible-light Heliospheric Imager (HI-1; covering the heliosphere between 15 to $84R_\odot$)[35].

From the Virtual Solar Observatory, we retrieved the EUVI (195 Å filter; Extended Data Fig. 6) and COR2 (double exposure total brightness) data covering the period from 2018-08-07 to 2018-08-12, at a cadence of 40 minutes. These data are processed using the standard SECCHI_PREP procedure that is available in SolarSoft. We extracted the K-corona signal from the COR2 images by subtracting pixel-wise minimum intensity from the entire time period considered (this will essentially remove the F-corona signal from the images). We retrieved the level-2 F-corona subtracted HI-1 images from the UK Solar System Data Centre (http://www.stereo.rl.ac.uk/). Besides coronal signal, HI-1 also images background star field. We suppressed the intensity of these background stars with peak intensities greater than $3\times10^{-13}$ the mean solar brightness, using standard procedures available in SolarSoft.

**STEREO observations during days leading up to the first perihelion of Parker Solar Probe.**
About 10 days before the first perihelion of Parker Solar Probe on 2018-11-06, the pair of coronal holes and the decaying active region appeared on the west-limb from the vantage point of STEREO-A, at Earth Ecliptic longitude of about -103°. We retrieved the EUVI, COR-2, and HI-1 data covering a five-day period from 2018-10-25 to 2018-10-29 from the Virtual Solar Observatory. These data are processed in a similar way as described in the previous paragraph (see Extended Data Fig. 6).

**STEREO composite maps.** We created composite STEREO images by projecting EUVI, COR2, HI-1 data on to a grid of conformal radial coordinates and position angle, using the World Coordinate System information available in the headers of respective FITS files. In Fig. 4 the EUVI images are plotted in log-scale and the COR-2 and HI-1 images are scaled by the cube of radial distance[56].

## Description of 3D MHD coronal model

The 3D MHD coronal simulation used here was produced with the MAS thermodynamic MHD global coronal model[27,57]. This model solves the resistive MHD equations on a global, non-uniform spherical mesh from 1–30$R_\odot$. MAS also solves for non-ideal energy transport terms relevant to low-coronal hydrodynamics (coronal heating, radiative losses, electron heat conduction) as well as auxiliary equations that represent the propagation, reflection, and dissipation of low-frequency Alfvénic fluctuations. Together these form the MAS Wave-Turbulence-Driven (MAS-WTD) model for coronal heating[27,58,59]. For our purposes, the MAS-WTD model provides a volume-filling 3D magnetic field and plasma state that includes the balance between magnetic and hydrodynamic forces. Most importantly, this computation is not restricted to assumptions about the plasma beta or source-surface radius, which may be essential for understanding dynamics in the middle corona. Combined with forward modeling (Fig. 3c) and comparisons to SDO/AIA data (Extended Data Fig. 5), we use this simulation to reasonably assert that our physical inferences and magnetic field analysis, including the heliospheric current sheet (HCS) location, AR cusp formation height, and coronal hole properties (Fig. 3 and Extended Data Fig. 4) are relevant to the observations at hand.

The most important inputs to the MAS-WTD model are the magnetic boundary conditions at the surface, the computational mesh, and the coronal heating model. The magnetic boundary conditions are based on the full-sun SDO/HMI synoptic map for CR2207, which represents radial magnetic fields measured at the solar surface as derived from line-of-sight magnetograms spanning 2018-08-06 to 2018-09-02. This map is then filled using a custom pole-filling procedure to add high-resolution flux-patches at the poles while matching the net flux in this region as derived from long-term observations[27]. The map is then multiplied by 1.4 (to roughly match HMI vector B observations for which the MAS-WTD model was calibrated) and re-gridded using a flux-preserving algorithm to the final MAS mesh. For this case, we use a relatively high-resolution mesh (364×344×504 in $r, \theta, \phi$) with a smoothly varying but non-uniform angular resolution that focuses the majority of mesh points around the AR and CH complex that is the subject of this study ($\Delta\phi$ = 0.005 rad here, which transitions to $\Delta\phi$ = 0.03 rad on the opposite side of the Sun). The next step is to smooth the re-gridded map to match the mesh resolution. Because of the relatively weak fields on the Sun at this time, we experimented with using less smoothing than previous simulations. This helps bring out the high-resolution complexity in the plasma state and magnetic field mapping by preserving the parasitic polarities of small-scale network flux and ephemeral regions (Extended Data Fig. 4). Lastly, for the heating model, we employ the same basic WTD approach and formulation as in ref.[27], but with updated parameters from the 2019 MAS-WTD eclipse prediction. This newer parametrization improves the electron density distribution in the middle corona as inferred by a comparison to white light observables[60].

The forward modeled images and comparisons to SDO/AIA imaging data in Extended Data Fig. 5 are computed using the SDO/AIA v10 effective areas and the CHIANTI 8.0.2 spectral synthesis package[61,62]. Observed emission in the AIA 171 Å channel, particularly from the quiet Sun and coronal hole regions, is likely dominated by small-scale transition region heating processes at the base of the solar corona[63], which could be attributed to radiation from plasma below 1 MK. Our MHD model does not capture these detailed small-scale heating events and the associated transition region emission. This is the reason why there are differences in the observed and synthesised AIA 171 Å emission maps in the Extended Data Fig. 5 (particularly in the coronal hole regions). This limitation, however, does not affect our inference and interpretation of the global magnetic topology.

The forward modeled composite in Fig. 3c uses a combination of forward modeled SDO/AIA 193Å emission and Thompson scattered polarized brightness, $pB$ (ref.[64]). Before compositing, each image is radially filtered and lightly unsharp masked at a fixed spatial scale. The radial filtering is done by dividing the raw observable at each pixel by a 1D monotonic profile in radius multiplied by a radial power law. The profile for each observable is determined simply by forward modeling the same observable using 1D radial profiles of the average temperatures and densities from a related simulation. The 3D volumetric renderings of the squashing factor[28] in Figs. 3a-b are based on a volumetric 3D magnetic field mapping done at twice the simulation resolution in each direction (8× overall). The rendered images are produced in the same manner as described in ref.[27].

**Data availability.** SUVI data (2018-08-07 to 2018-09-13) are available at ref.[65]. SDO full disk data (2018-08-08 to 2018-09-09) are available at Joint Science Operations Center (AIA's 171 Å, 193 Å, and 211 Å filter maps at http://jsoc.stanford.edu/AIA/AIA_lev1.html and HMI's line-of-sight magnetic field data at http://jsoc.stanford.edu/HMI/Magnetograms.html). SDO/AIA synoptic maps are available at http://satdat.oulu.fi/solar_data/. SDO/HMI synoptic radial magnetic field maps are available at http://hmi.stanford.edu/QMap/. LASCO-C2 data (2018-08-07 to 2018-09-13) and STEREO's EUVI-195 Å and COR-2 data (2018-08-07 to 2018-08-12 and 2018-10-25 to 2018-10-29) are available via Virtual Solar Observatory (https://sdac.virtualsolar.org/cgi/search). STEREO's HI-1 data (2018-08-07 to 2018-08-12 and 2018-10-25 to 2018-10-29) are available via the STEREO Archive of the UK Solar System Data Centre (https://www.ukssdc.ac.uk/solar/stereo/data.html). The LASCO SUVI merged FITS files are archived at ref.[66].

**Code availability.** PFSS code is available via SolarSoftWare. NASA's Community Coordinated Modeling Center (CCMC) at https://ccmc.gsfc.nasa.gov has a version of the MAS code for producing simulation runs on demand.

**Acknowledgments.** We thank GOES, LASCO, SDO, and STEREO teams for making the data publicly available. We thank Richard Harrison and Jackie Davies (both at STFC - UKRI) for their help with STEREO HI-1 data. CHIANTI is a collaborative project involving George Mason University, the University of Michigan (USA), University of Cambridge (UK) and NASA Goddard Space Flight Center (USA). L.P.C was supported by funding from the European Research Council (ERC) under the European Union's Horizon 2020 research and innovation programme (grant agreement No 695075; PI: Sami K. Solanki, MPS). D.B.S. acknowledges support for GOES-R activities at CIRES via National Oceanic and Atmospheric Administration cooperative agreement NA17OAR4320101 and NASA's HGI program, grant 80NSSC22K0523. C.D. acknowledges support from the NASA HSR, LWS, and HSOC programs (grants 80NSSC18K1129, 80NSSC20K0192, and 80NSSC20K1285).

**Author contributions.** L.P.C., D.B.S., and C.E.D. contributed to observational analysis. C.D. computed 3D MHD coronal models. A.K.H provided context to solar wind models. All authors contributed to the manuscript and discussed the results.

**Competing interests.** There are no competing interests.

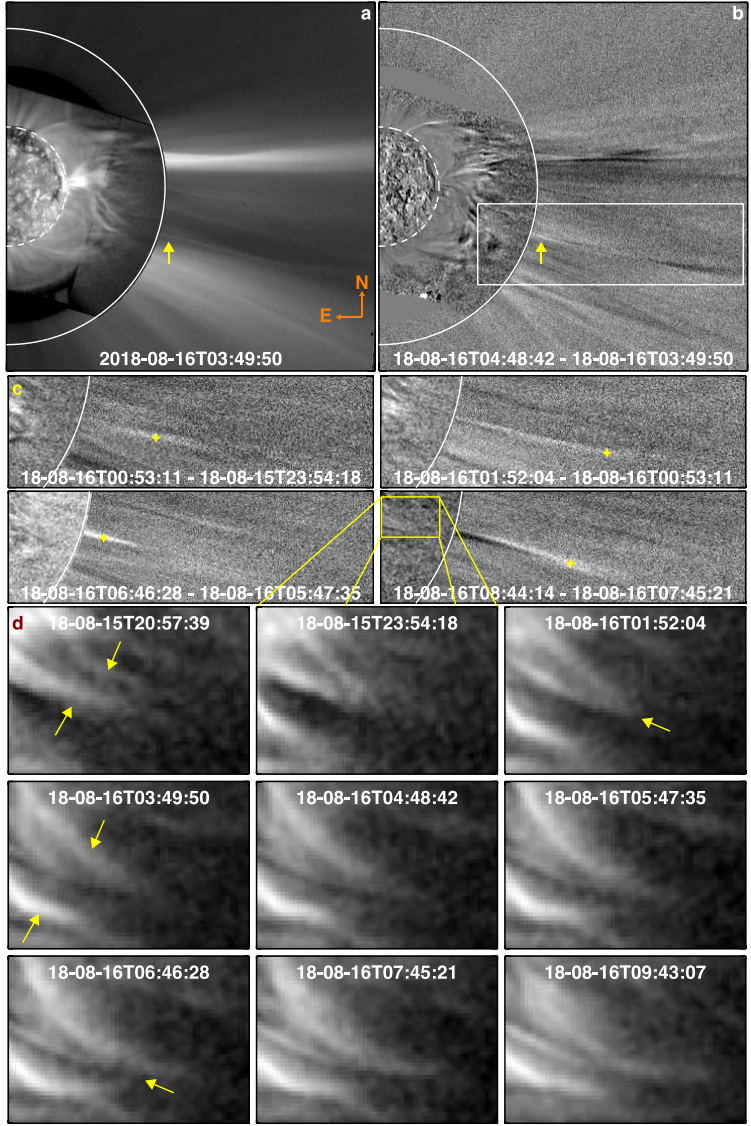

**Fig. 1: Magnetic driver of a slow solar wind stream. a,** SUVI 195 Å and LASCO composite image showing the corona over the west limb. The arrow points to an outflowing solar wind stream. The dashed and solid arcs are at one and 2.65 solar radii from the Sun center. **b,** Difference image obtained by subtracting two composite images with timestamps as indicated is shown. **c,** The two rows show a sequence of difference images in a field of view covered by the white rectangle in panel **b**. The star symbols guide the eye to follow a solar wind stream. **d,** The three rows show a time sequence of SUVI images (field of view marked by a box in panel **c**). The slanted arrows point to interacting structures in the middle corona. See Supplementary Video 1.

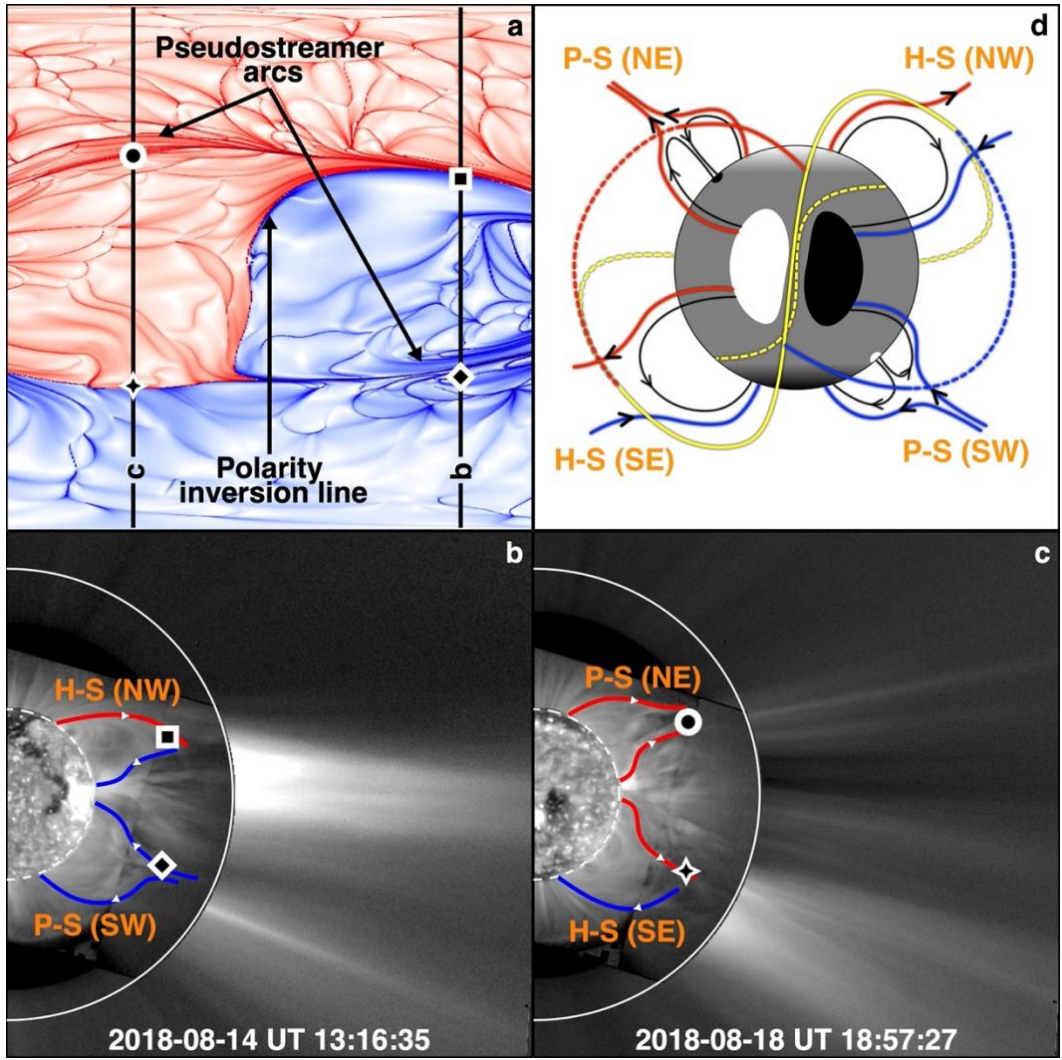

**Fig. 2: Imprints of S-web in the observed coronal web. a,** Synoptic signed log$Q$ map at 3R$_\odot$, spanning the CH-AR system (field of view marked in Extended Data Figs. 3 and 4). Red(blue) shaded regions represent outward(inward) oriented magnetic field. The arrows point to the prominent S-web features. The four different symbols along the vertical lines (marking the Carrington longitudes at the west-limb) identify the S-web features that are labeled in panels **b** and **c**. **b,** Solar west-limb map along line-b in panel **a**. Overlaid red(blue) curves represent outward(inward) magnetic structures at those locations, outlining the magnetic topology of helmet streamers (H-S) and pseudostreamers (P-S) in plane of sky. **c,** Solar west-limb map along line-c in panel **a** in the same style as panel **b**. **d,** A bird's eye view illustration of the H-S and P-S features. The polarity inversion line (yellow solid and dashed curves) and prominent pseudostreamers arcs (red and blue solid and dashed curves) are marked. The observed coronal web is bounded by these topological structures. The lighter and darker shaded regions on the grey solar surface represent magnetic polarities (exaggerated for clarity). See Supplementary Video 5.

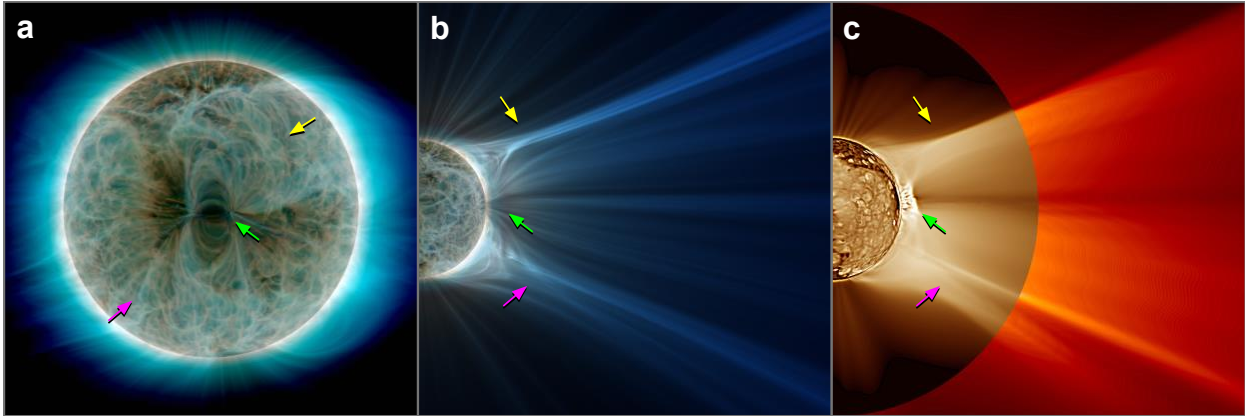

**Fig. 3: Global magnetic skeleton and coronal structures based on 3D MHD simulations. a,** 3D volume render of squashing factor ($Q$) from Earth's perspective on 2018-08-10 UT 14:00. FOV is ±1.4$R_\odot$ (see Supplementary Video 7). **b,** 3D $Q$ volume render for Earth's perspective on 2018-08-17 UT 02:00. FOV is 0–6$R_\odot$ in $X$, ±3$R_\odot$ in $Y$ (see Supplementary Video 8). **c,** Forward modeled AIA 193 Å (brown) and polarized brightness (red) unsharp mask filtered composite for the same FOV and view as in **b**. The yellow and magenta arrows point to the northeast and southwest streamers. The green arrow points to the top or cusp of closed coronal loops.

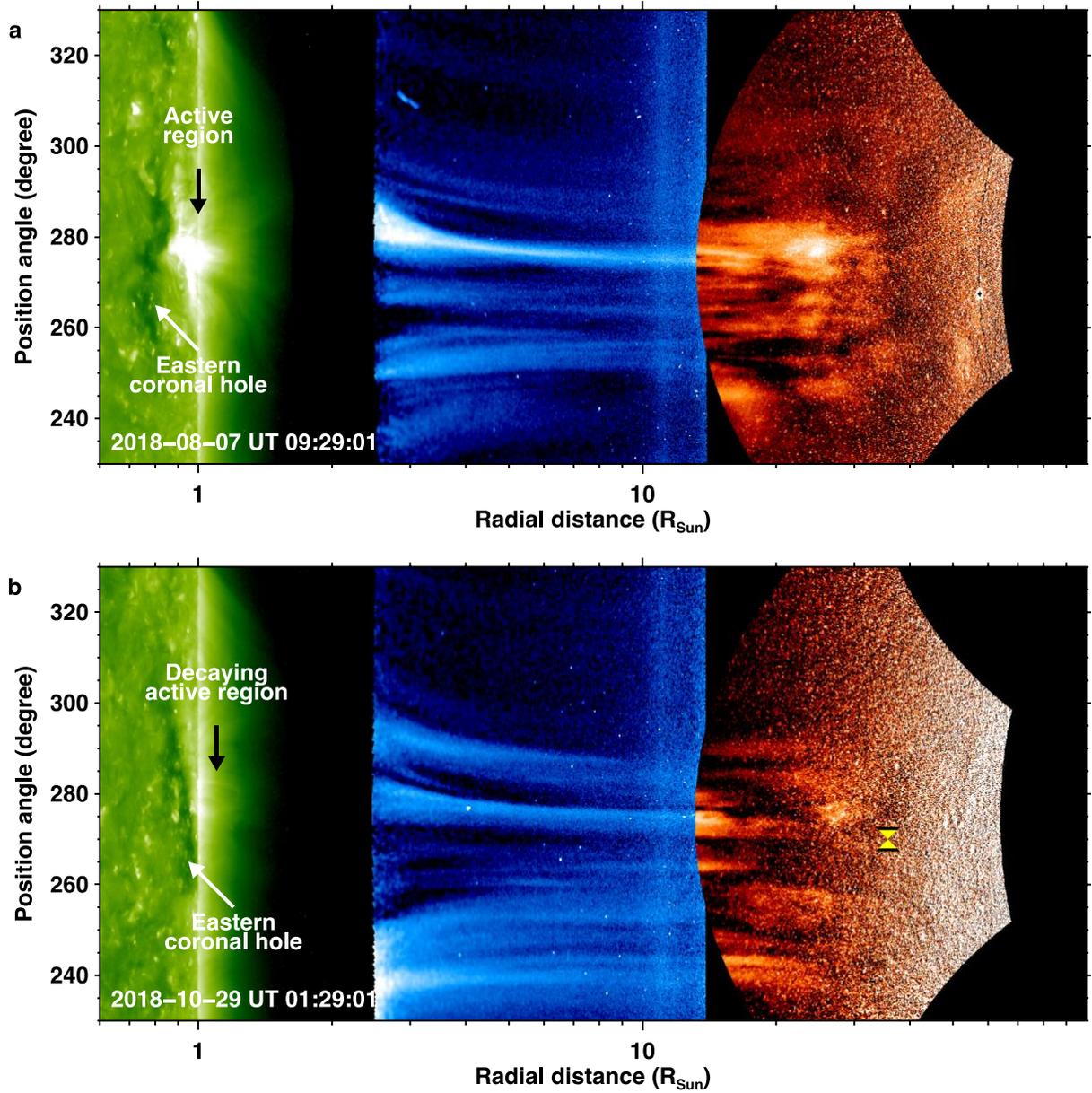

**Fig. 4: Heliospheric connection.** Solar wind structures seen in the outer corona (COR-2; blue) and heliosphere (HI-1; red) and their association with low-latitude CH-AR system (EUVI; green; see Extended data Fig. 6). **a,** Closer in time to SUVI observations (see Supplementary Video 9). **b,** Closer in time to the first perihelion of PSP (the symbol marks the approximate plane-of-sky position of PSP during its perihelion on 2018-11-06; see Supplementary Video 10). The position angle increases in the anti-clockwise direction from the solar north.

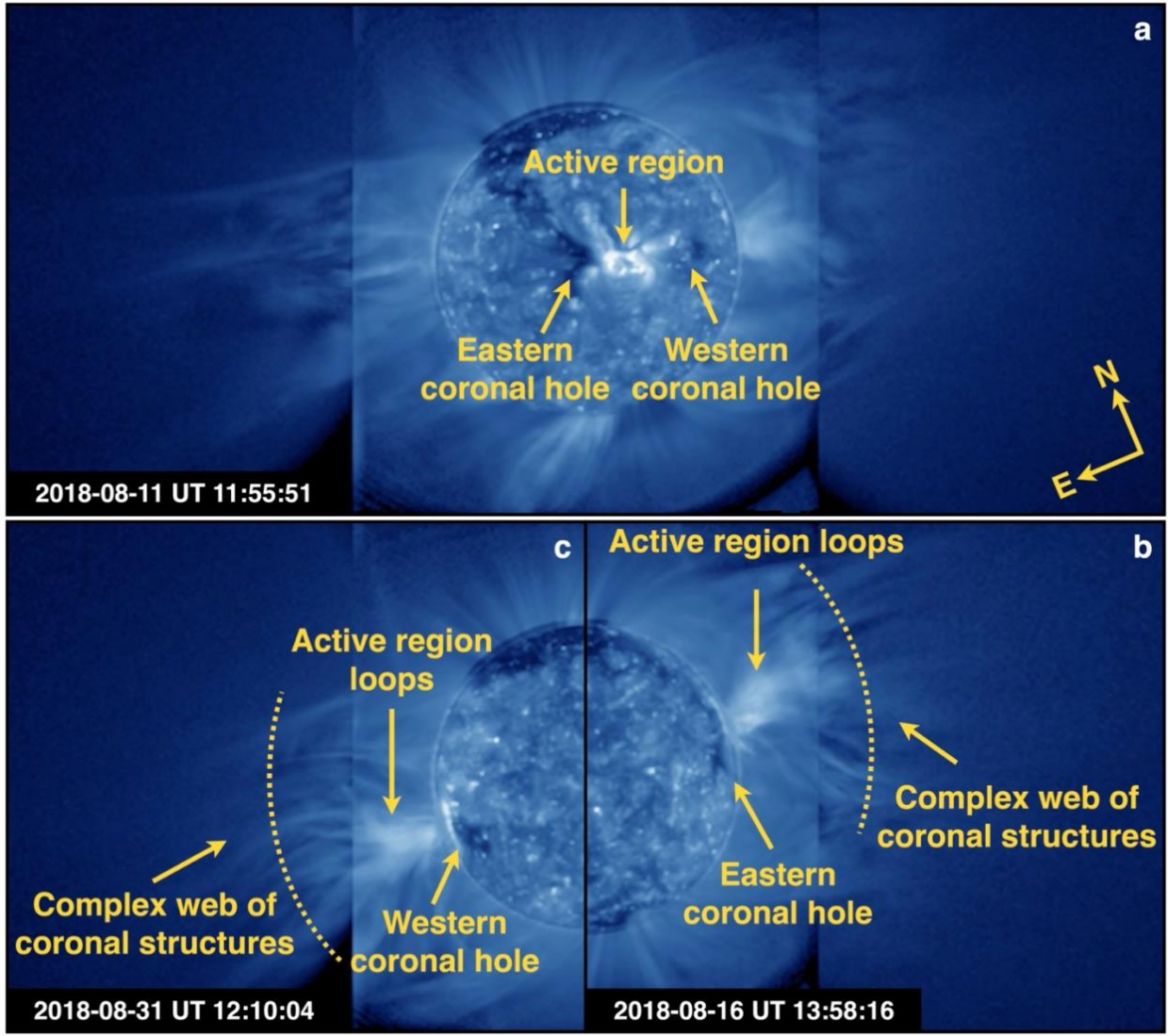

**Extended Data Fig. 1: Overview of SUVI observations. a,** Coronal hole pair (identified as eastern and western) and the embedded active region observed on 2018-08-11. **b,** Western hemisphere of the solar corona on 2018-08-16 showing the active region loops over the limb and the eastern coronal hole (the western coronal hole is behind the limb). **c,** Eastern hemisphere of the solar corona on 2018-08-31. Coronal loops over the limb from the same active region and the western coronal hole are visible (the eastern coronal hole is behind the limb). The dotted arcs in the lower panels highlight the latitudinally extended complex web of coronal structures observed over the CH-AR system. These observations are recorded by the SUVI 195 Å EUV filter.

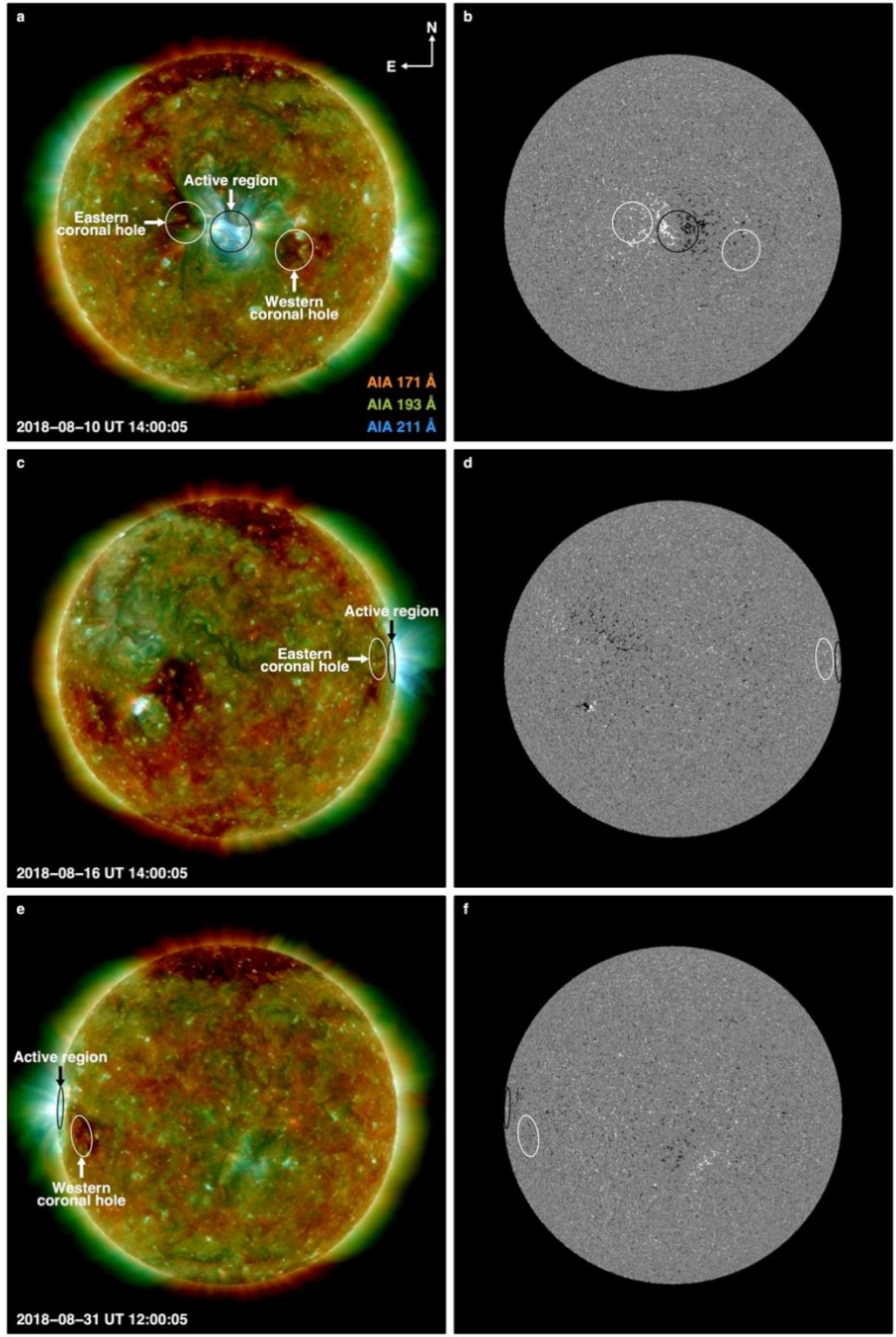

**Extended Data Fig. 2: The lower-coronal and surface magnetic field features associated with the coronal web. a, c,** and **e,** Three-filter composite image (AIA 171 Å: red; 193 Å: green; and 211 Å: blue). The circles outline the two coronal holes (white) and the embedded active region (black). **b, d,** and **f,** HMI photospheric line-of-sight magnetic field map (light shaded regions represent positive polarity magnetic fields, whereas the dark shaded regions represent negative polarity magnetic fields; map saturated at ±50 G).

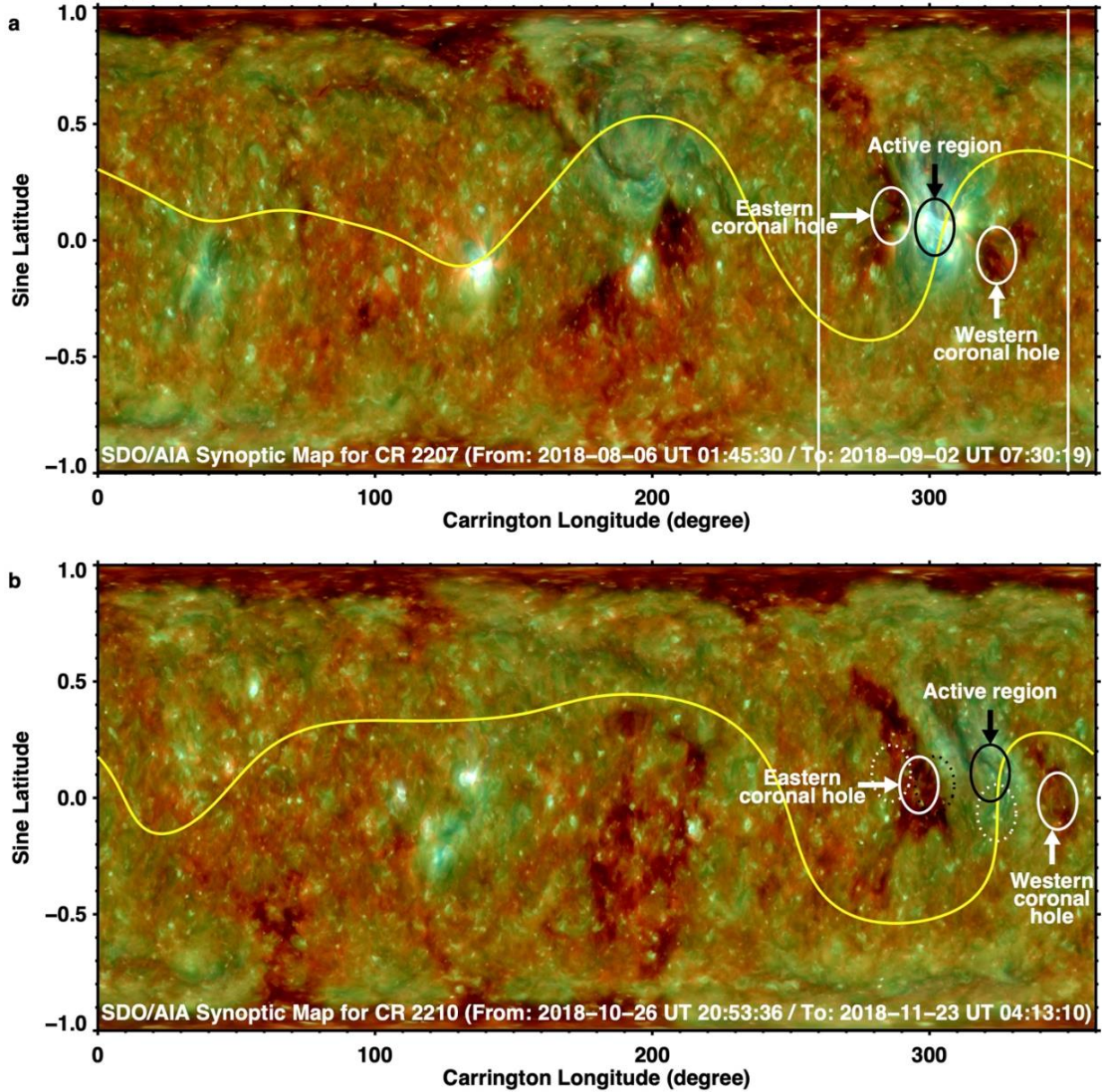

**Extended Data Fig. 3: The lower-coronal features associated with the coronal web. a,** SDO/AIA synoptic map for the Carrington rotation (CR) 2207 overlapping observing period of SUVI campaign (AIA 171 Å: red; 193 Å: green; and 211 Å: blue). Time increases with decreasing Carrington longitudes. The ellipses cover the same regions as in Extended Data Fig. 2a. The yellow curve is the polarity inversion at $2R_\odot$, based on a PFSS model. The white lines outline the extent of field of view displayed in Fig. 3a. **b,** Same format as a, but displayed for SDO/AIA synoptic maps covering CR2210. The dotted circles outline the same Carrington longitudes and latitudes of the pair of coronal holes and the embedded active region as determined from CR2207. The solid circles mark the new positions of these features as identified in CR2210.

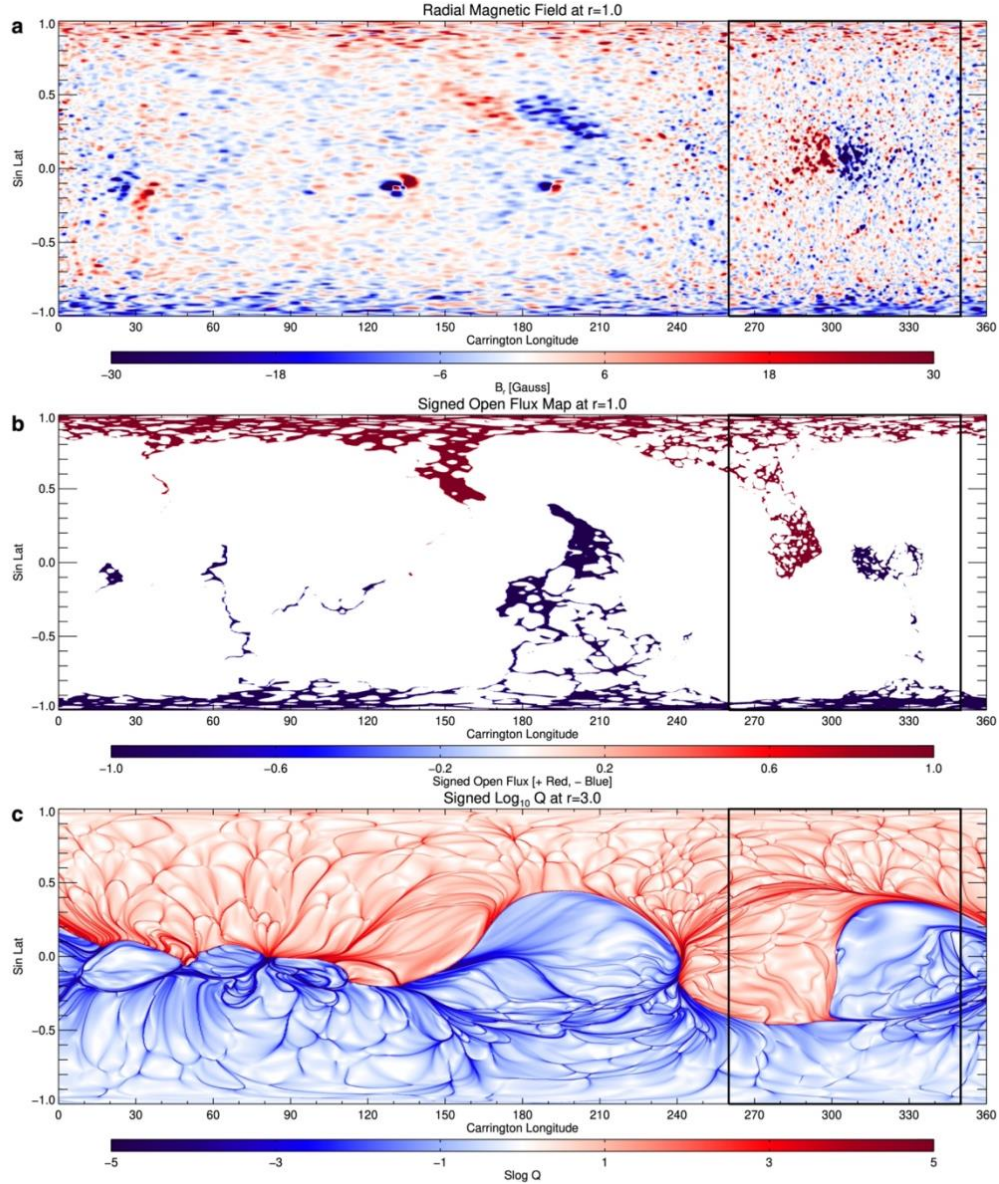

**Extended Data Fig. 4**: **Magnetic diagnostics. a,** Radial magnetic field at the inner boundary of the high-resolution MAS MHD simulations. This map is derived from the HMI synoptic map for CR2207 ('hmi.synoptic_mr_720s'). The black box covers the CH-AR system, and has a width of 90°, spanning from 260° to 350° Carrington longitudes (white box in Extended Data Fig. 3a). Red(blue) shaded regions represent radial outward(inward) magnetic fields. **b,** Synoptic map displaying regions of open magnetic flux that map coronal hole features. **c,** Synoptic signed logQ map at $3R_\odot$ for CR2207, with the mapping domain from 1 to $3R_\odot$, colored red(blue) by the outward(inward) orientation of magnetic field.

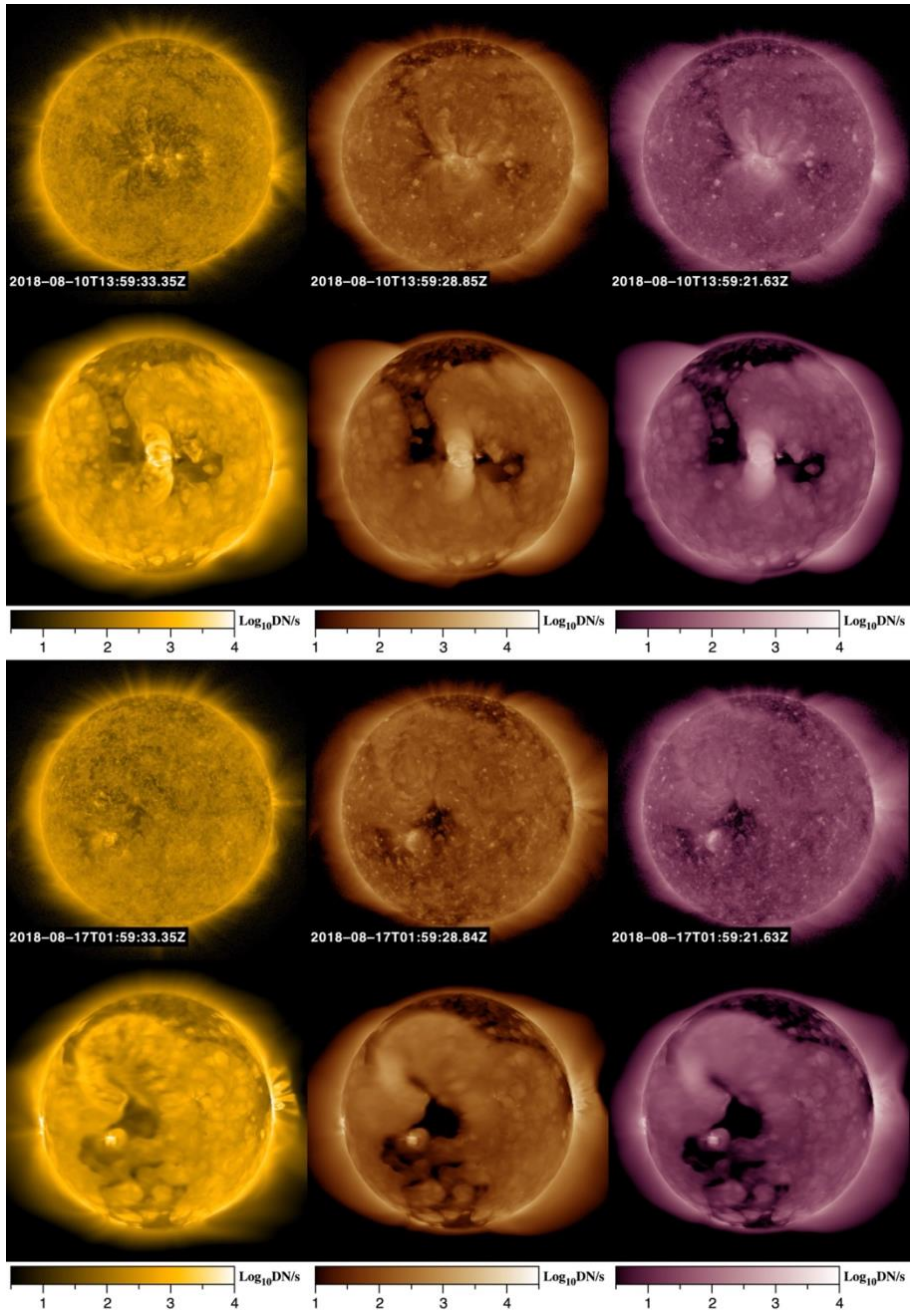

**Extended Data Fig. 5: Comparison of coronal observations with the MHD model.** In each segment, upper panels show SDO AIA coronal observations (AIA 171 Å: left; AIA 193 Å: middle; AIA 211 Å: right) and the lower panels show corresponding synthetic emission from the MHD model. SDO observations in the top segment are obtained on 2018-08-10 UT 14:00 when the CH-AR system is close to the disk center (see Extended Data Fig. 2a). SDO observations in the bottom segment are obtained on 2018-08-17 UT 02:00 when the CH-AR system is close to the west-limb (see Extended Data Fig. 2c).

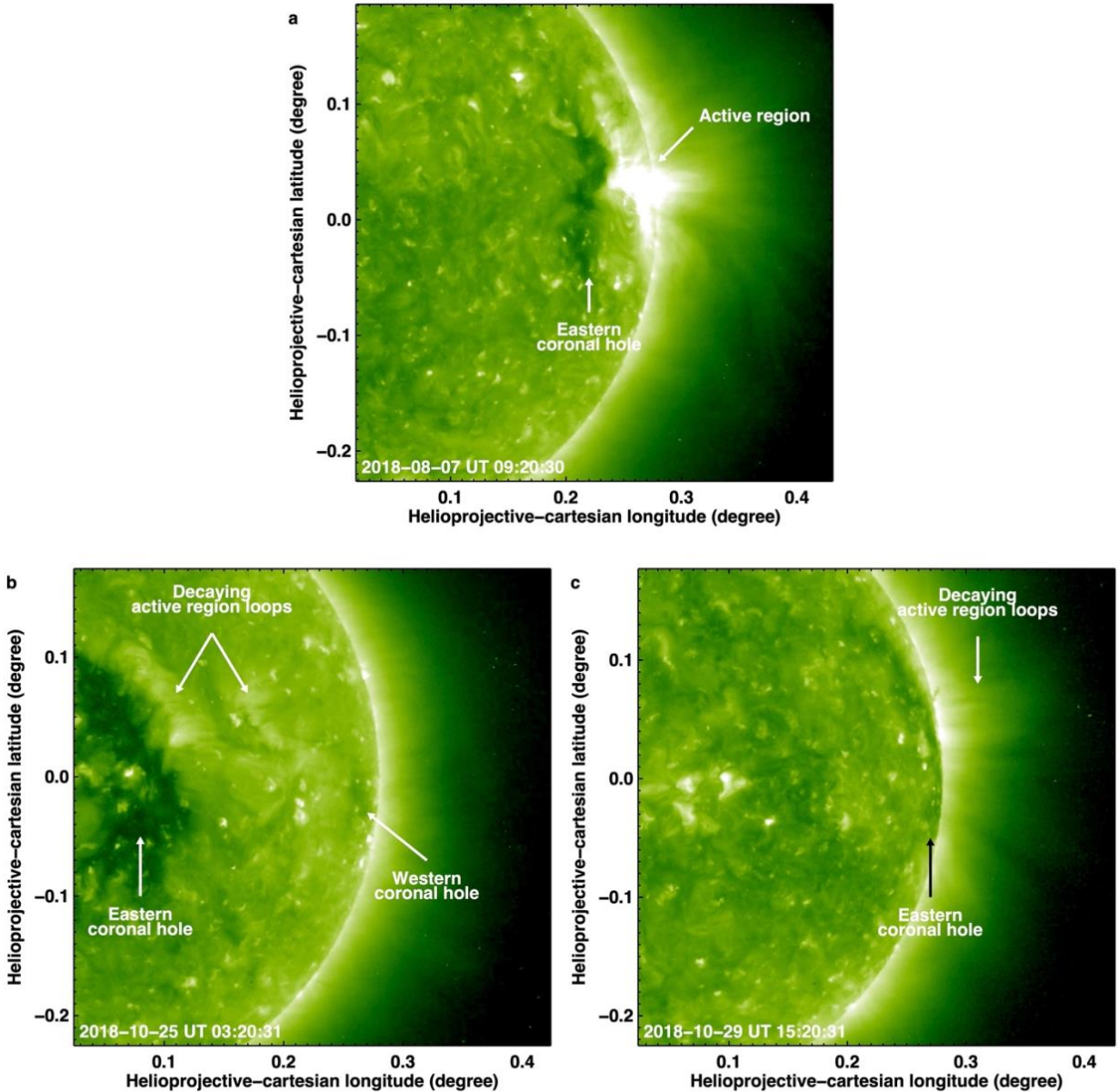

**Extended Data Fig. 6: The lower-coronal features associated with the coronal web.** Close-up of the lower-coronal features observed with STEREO/EUVI 195 Å filter. The eastern coronal hole and the active region at the limb are marked. a, Closer in time to SUVI observations. b and c, Closer in time to the first perihelion of PSP.

**This Supplementary Information file contains:**

- **Supplementary Notes**
    1. **Additional examples**
    2. **Topology of the observed coronal web and its relation to the S-web**
    3. **Dynamics and flow speeds of solar wind streams emerging from the S-web**

- **Supplementary Figures 1–6**
- **References**

# Supplementary Notes

## 1. Additional examples

In the main text we presented an example of a slow solar wind stream driven by a system of interacting magnetic structures in the middle corona in a complex coronal web formed over a low-latitude CH-AR system (Fig. 1). Here we discuss three additional examples that further illustrate a close association between dynamics in the complex coronal web and slow solar wind structures.

Within LASCO's field of view, we identified the ejection of a blob along a bright streamer (Supplementary Figure 1a-c). We observed that the streamer blob is ejected outward when a pair of initially open EUV structures in the middle coronal web approach each other and reconnect at an X-type configuration to form a closed loop (Supplementary Figure 1d). Similarly, in Supplementary Figure 2, we present the example of a narrow wind stream originating at the tip of interacting and reconnecting structures in the complex coronal web. The example in Supplementary Figure 3 is a case when the complex coronal web over the CH-AR system appeared rotated to the east-limb (see also Extended Data Fig. 1c. In this example, in addition to the detection of solar wind stream being ejected from the tips of a pair of interacting and reconnecting structures in the coronal web (Supplementary Figure 3a-b), we also detected apparent sunward flows from the site of reconnection (slanted arrows in Supplementary Figure 3 indicate the direction of flows). This system of bidirectional flows (i.e., outward solar wind stream and a corresponding sunward flow) provides further evidence that the complex coronal separatrix web is dynamically evolving through reconnection and driving solar wind streams.

## 2. Topology of the observed coronal web and its relation to the S-web

Coronal holes, which quite commonly form in the unipolar remnants of active regions confined to low-latitudes, can reshape the streamer belt, extending it to high latitudes by distorting the central heliospheric current sheet (HCS)[9]. To this end, to gain insights into the topology of the coronal web that is associated with a pair of coronal holes and a low-latitude active region, we examined low-coronal EUV images and photospheric magnetic field data, respectively from the Atmospheric Imaging Assembly (AIA) and Helioseismic and Magnetic Imager (HMI) instruments on board the Solar Dynamics Observatory (SDO) (Methods). Based on HMI magnetic field data, we found that the embedded active region was decaying during the course of SUVI observations, and did not exhibit any systematic large-scale emergence of magnetic flux. Thus, the very existence of the complex coronal web in itself is not associated with any explosive activity of the decaying active region (e.g., flares). Next, based on potential field source surface (PFSS) extrapolations, we found that the global polarity inversion line (that separates the radial positive and negative polarity magnetic fields) is warped and bisects the active region with a coronal hole on either side (Extended Data Fig. 3). Since this global polarity inversion line is a true separatrix[9], it forms HCS. In this case, the HCS is approximately north-south aligned, passing in between the two coronal holes. When projected in the plane of the sky, this would take the form of a north-south aligned, in-plane current sheet. This suggests that the coronal web is associated with the north-south aligned current sheet, that is governed by the global distribution of magnetic fields.

The coronal web is clearly observed as a latitudinally extended complex of structures in the middle corona over the east-west aligned CH-AR system (main text; see Extended Data Fig. 1). Here we elaborate on the morphology of the observed coronal web and its direct relation to the S-web (consisting quasi-separatrix layers and separatrices)[9,18].

The observed 2D morphology of the coronal web in the image plane is governed by the global distribution of magnetic fields. As mentioned in Methods, the active region is composed of leading negative polarity magnetic fields and trailing positive polarity magnetic fields. The orientation of the dipolar magnetic field of the Sun is such that the magnetic north(south) pole is along the Sun's true north(south) pole. The global polarity inversion line (PIL), orients itself along the north-south direction in the lower latitudes, closer to the Sun's equator, as it bisects the active region. Then it warps into an east-west aligned PIL at latitudes above ~ 20° in both hemispheres. This warped PIL (bisecting the active region) is between Carrington longitudes of ~ 260°–350° (Extended Data Fig. 3a). Thus, the HCS overlying this warped PIL is also warped. This north-south extent of the warped HCS governs the latitudinal extent of the observed coronal web.

Based on the warping of the HCS, we expect a bipolar helmet streamer at the north-west of the active region, due to magnetic structures that cross the east-west PIL around the Carrington longitude of 330°. Because the direction of magnetic field changes perpendicular to the length of the streamer, it is embedded with a current sheet[9]. This north-west helmet streamer would then be a result of magnetic connections between the Sun's north pole (hosting positive polarity magnetic fields) and the leading negative polarity magnetic fields of the active region and the associated coronal hole. Similarly, we expect another bipolar helmet streamer south-east of the active region. This helmet streamer would then be a result of magnetic connections between the Sun's south pole (hosting negative polarity magnetic fields) and the trailing positive polarity magnetic fields of the active region and the associated coronal hole. In the Carrington synoptic map displayed in Extended Data Fig. 3a, time increases from right to left (i.e., with decreasing Carrington longitudes). As the Sun rotates, we first expect to observe the helmet streamer in the northern hemisphere (that is north-west of the active region) in the middle corona, projected over the west limb. As this feature rotates away from view, we then expect to observe the second helmet streamer in the southern hemisphere (that is south-east of the active region). Indeed, SUVI's middle coronal observations clearly revealed north-west and south-east helmet streamers, in that order (Fig. 2).

In addition to the bipolar helmet streamers, pseudostreamers, unipolar streamer stalks that are associated with distinct like-polarity magnetic features, form adjacent to the warped HCS. Because the direction of magnetic field remains the same perpendicular to the length of the pseudostreamer stalks, these features are not embedded with a current sheet, but host a magnetic null point[9]. Because of this morphological difference, pseudostreamers are clearly distinguishable from bipolar helmet streamers. The footprints of these pseudostreamers are seen as regions of strong gradients across like-polarity magnetic domains in the Synoptic signed log $Q$ map (Fig. 2a). Thus, the unipolar pseudostreamers associated with the CH-AR system are expected to appear at the north-east and south-west of the active region. Again, because of rotation of the system across the limb, the south-west pseudostreamer appears first and then the north-east one (Fig. 2a). Similar to the ordering of helmet streamers, SUVI did clearly observe both the south-west and north-east pseudostreamers, in that order (Fig. 2). In summary, SUVI first observed a northern helmet streamer and southern pseudostreamer pair, then nearly after five days of solar rotation, a northern pseudostreamer and a southern helmet streamer pair, over the west limb (Fig. 2). This same sequence repeats itself when the CH-AR system crosses the east limb, over a period of roughly five days, after rotating through the far side (Supplementary Figure 5).

On both these occasions (crossing of CH-AR system across west and east limb), SUVI observations revealed the radially protruding coronal web features (marked in Extended Data Fig. 1) that exhibited persistent dynamic evolution.

## 3. Dynamics and flow speeds of solar wind streams emerging from the S-web

Low in the corona (below ~ $1.5R_\odot$), we do not find clear differences in the loop structures associated with streamers and pseudostreamers. Above these altitudes the global magnetic topology begins to be more apparent. This is reflected in the clear morphological differences between streamers and pseudostreamers. These morphological differences also contribute to the differences in their dynamics in the middle corona. In streamers we do observe more signatures of retracting loops with emergence of blobs while the pseudostreamers are comparatively quiescent. The differences between the two become more apparent in the LASCO-C2 field of view. The streamers constitute more blob-like ejections while the pseudostreamers are composed of ray-like wind streams.

Speeds of outflowing slow solar wind streams have been studied using LASCO-C2 and LASCO-C3 data[23,24]. Within the LASCO-C2 field of view that extends roughly to $6R_\odot$, these observations show that speeds of slow wind streams range from a few 10 km s$^{-1}$ to roughly 200 km s$^{-1}$. The speeds continue to increase to 300 km s$^{-1}$ to 400 km s$^{-1}$ at $30R_\odot$ (i.e., around the outer edge of LASCO-C3 field of view).

For the four examples of wind streams presented in Figure 1 and Supplementary Figures 1–3, we track the features from the inner to outer edge of the LASCO-C2 field of view to investigate their flow speeds. To this end, we created distance-time maps of each of the wind streams by placing an artificial nine-pixels-wide slit along the path of that wind stream. In these maps, the wind streams are seen as slanted ridges, with their slopes indicating the speed of propagation (Supplementary Figure 4). In the examples we considered, we find that the speeds range from 35 km s$^{-1}$ to 160 km s$^{-1}$. These speeds are broadly consistent with slow wind speeds in the literature[23,24]. From Supplementary Figure 4, we note that the example of pseudostreamer wind stream has higher speed (panel a) compared to the speeds of streams emerging from the HCS (panels b-d), which is consistent with earlier studies[67].

We retrieve relevant plasma parameters (electron number density and velocity) from the MHD model for a more quantitative comparison with observations. In the upper two panels of Supplementary Figure 6 we show the electron number density and radial velocity from the MHD model sliced at $3R_\odot$. At the large-scale HCS crossing identified in the signed $\log Q$ mapping, we observe the largest densities (Supplementary Figure 6a). This high density HCS crossing from the model corresponds to the brighter helmet streamers in our observations. These high-density streams over the HCS are generally anti-correlated with the radial velocity (Supplementary Figure 6b). In the model, the large-scale pseudostreamer arcs (see Fig. 2a) are seen as relatively lower density features compared to the HCS structure. Our model predicts relatively higher speeds from the pseudostreamers, which is consistent with our observations. These are the typical characteristics expected for the slow solar wind. Away from the boundaries of these large-scale topological features, we also see cellular pockets of smaller but correlated density and velocity contrasts. These align well with the features in the signed $\log Q$ mapping, which are formed by separatrices in the mapping due to presence of distinct small-scale flux concentrations at the surface within the larger open field region. These large- and small-scale links between signed $\log Q$ and the MHD quantities again illustrate how the geometry of the magnetic field and complexity in its mapping may structure the solar wind.

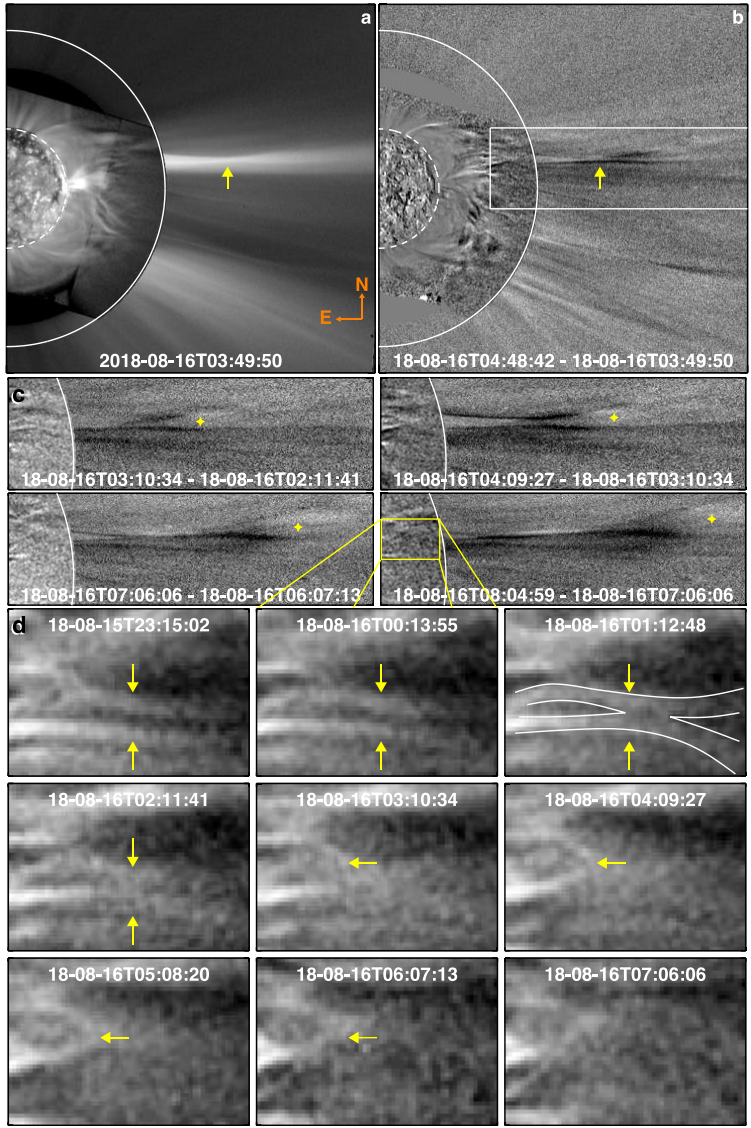

**Supplementary Figure 1: Additional example-1 pointing to the magnetic driver of a slow solar wind stream.** The format is same as in Fig. 1 in the main text. In **d,** the vertical arrows point to a pair of interacting (locally) open coronal structures in the middle coronal web (curves outline an X-type configuration). The horizontal arrows in subsequent panels point to a closed loop that formed as a result of reconnection between the interacting pair. See Supplementary Video 2.

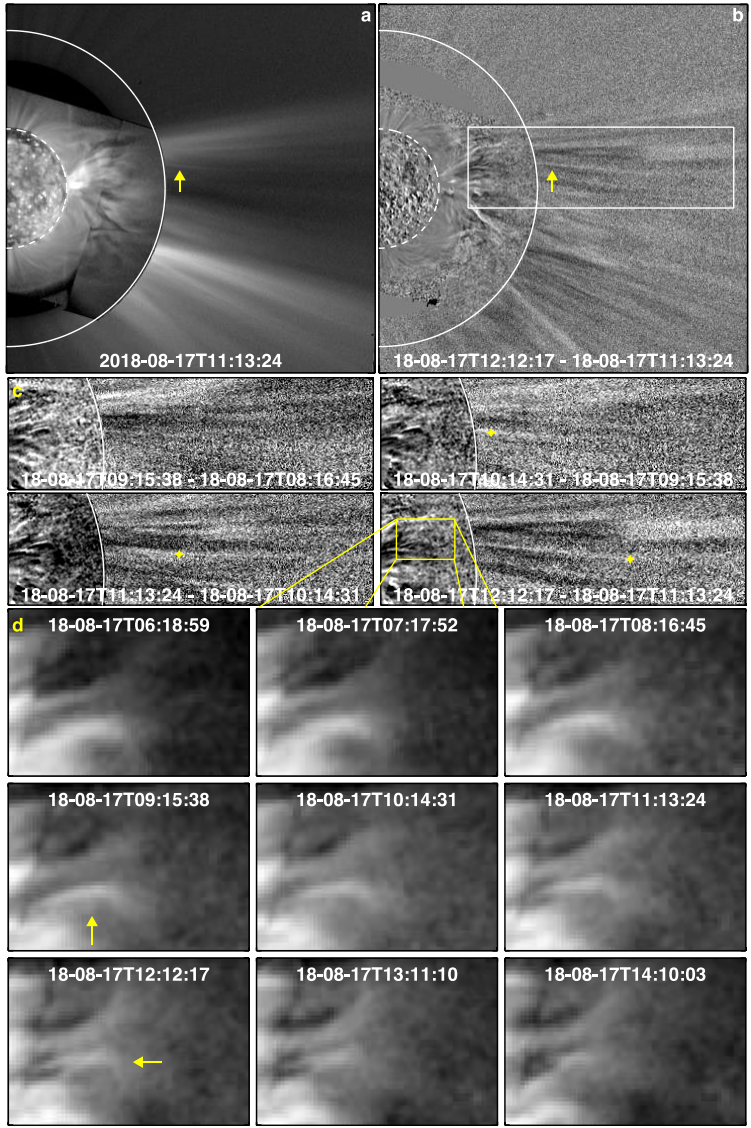

**Supplementary Figure 2: Additional example-2 pointing to the magnetic driver of a slow solar wind stream.** The format is same as in Fig. 1 in the main text. See Supplementary Video 3.

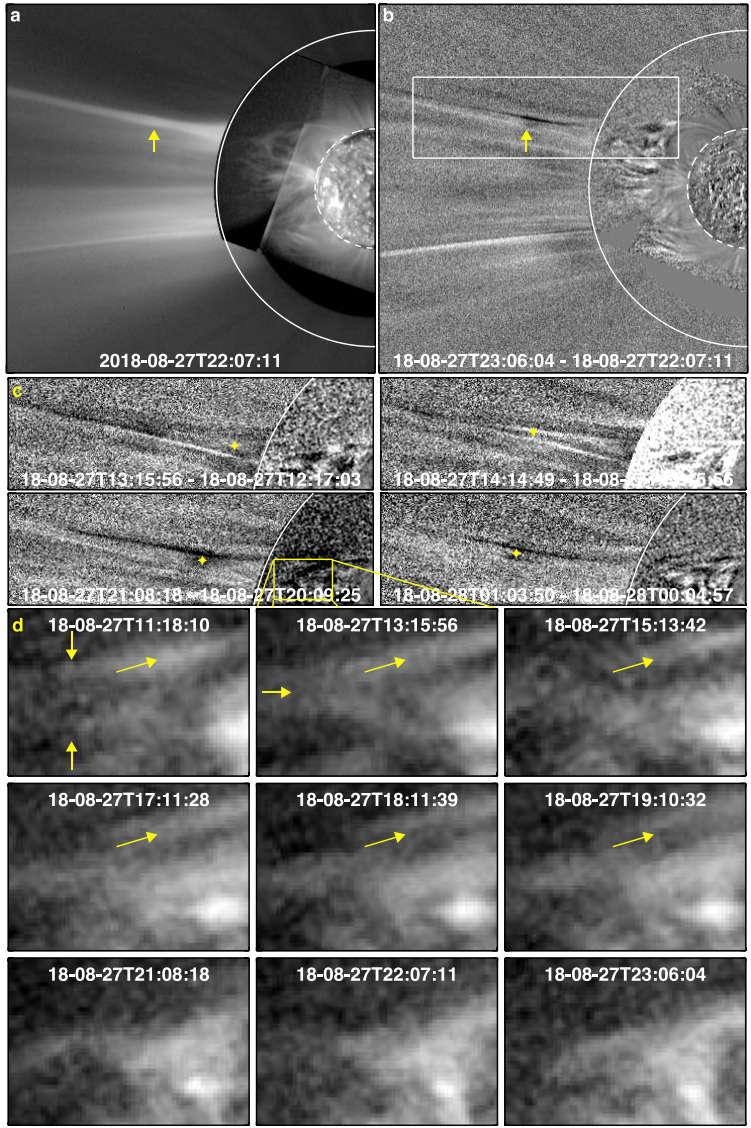

**Supplementary Figure 3: Additional example-3 pointing to the magnetic driver of a slow solar wind stream.** The format is same as in Fig. 1 in the main text. In this case, the CH-AR system is on the east-limb (see Extended Data Fig. 1). The slanted arrows in **d** point to the coronal material draining toward the Sun. See Supplementary Video 4.

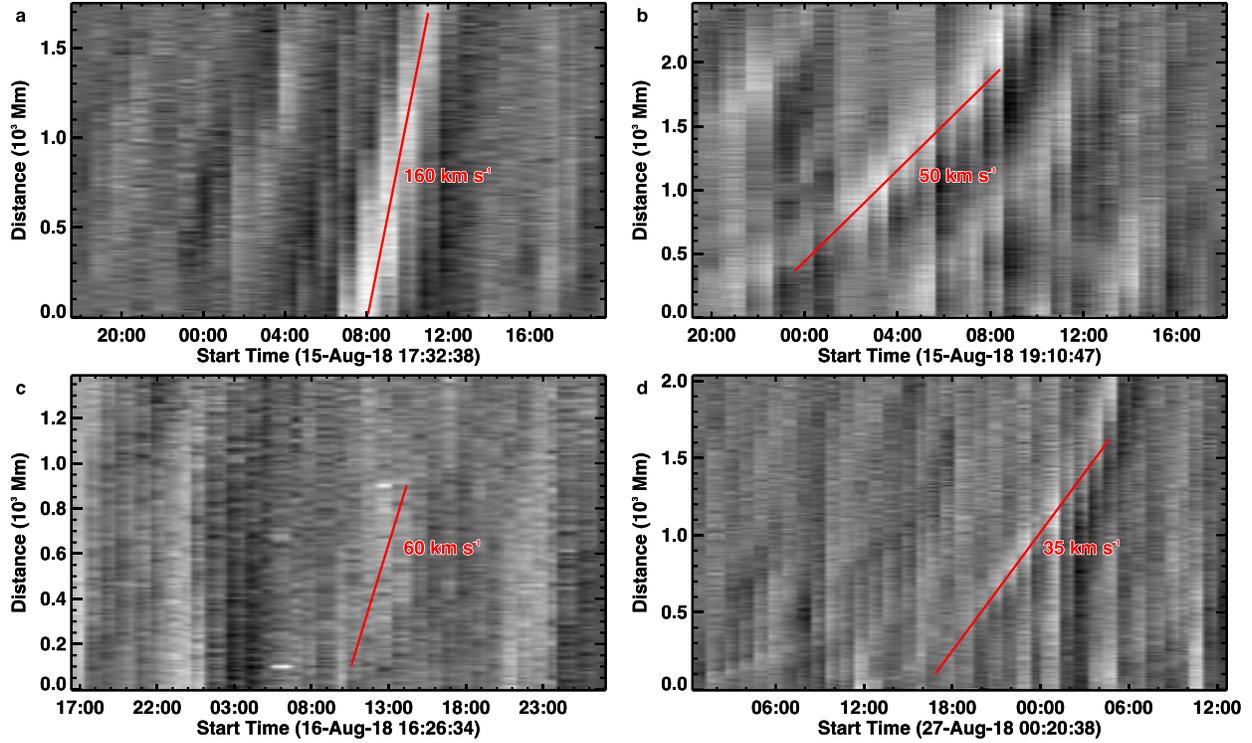

**Supplementary Figure 4: Outflow speeds of slow wind streams.** The distance-time stack plots of slow wind streams discussed in Fig. 1 and Supplementary Figures 1–3 are displayed in panels a-d, respectively. The outward propagating slow wind stream is seen as a bright slated ridge in these stack plots. The slanted fiducial line is overlaid on the respective ridges to guide the eye. The slope of this line in units of km s$^{-1}$ is also indicated. The 0 Mm point in these plots roughly corresponds to the inner edge of the LASCO field of view. The distance increases away from the Sun along the track of the solar wind stream.

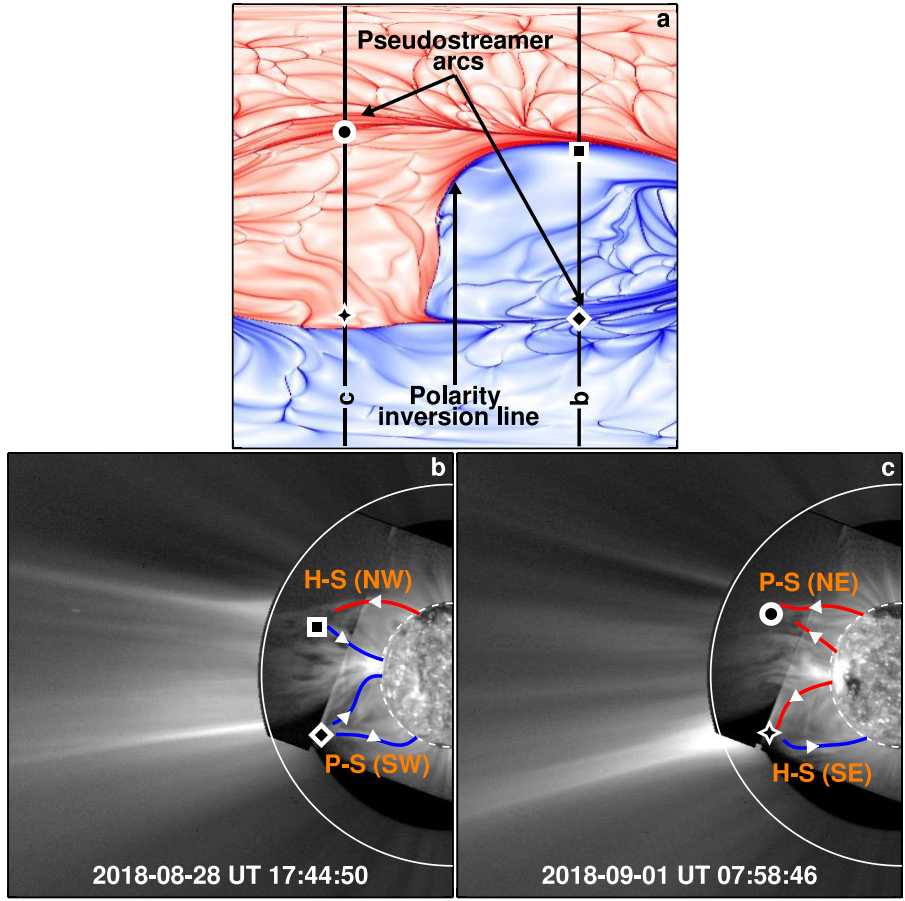

**Supplementary Figure 5: Imprints of S-web in the observed coronal web.** In the same format as Fig. 3a-c, but during the period when the CH-AR system crossed the east-limb. See Supplementary Video 6.

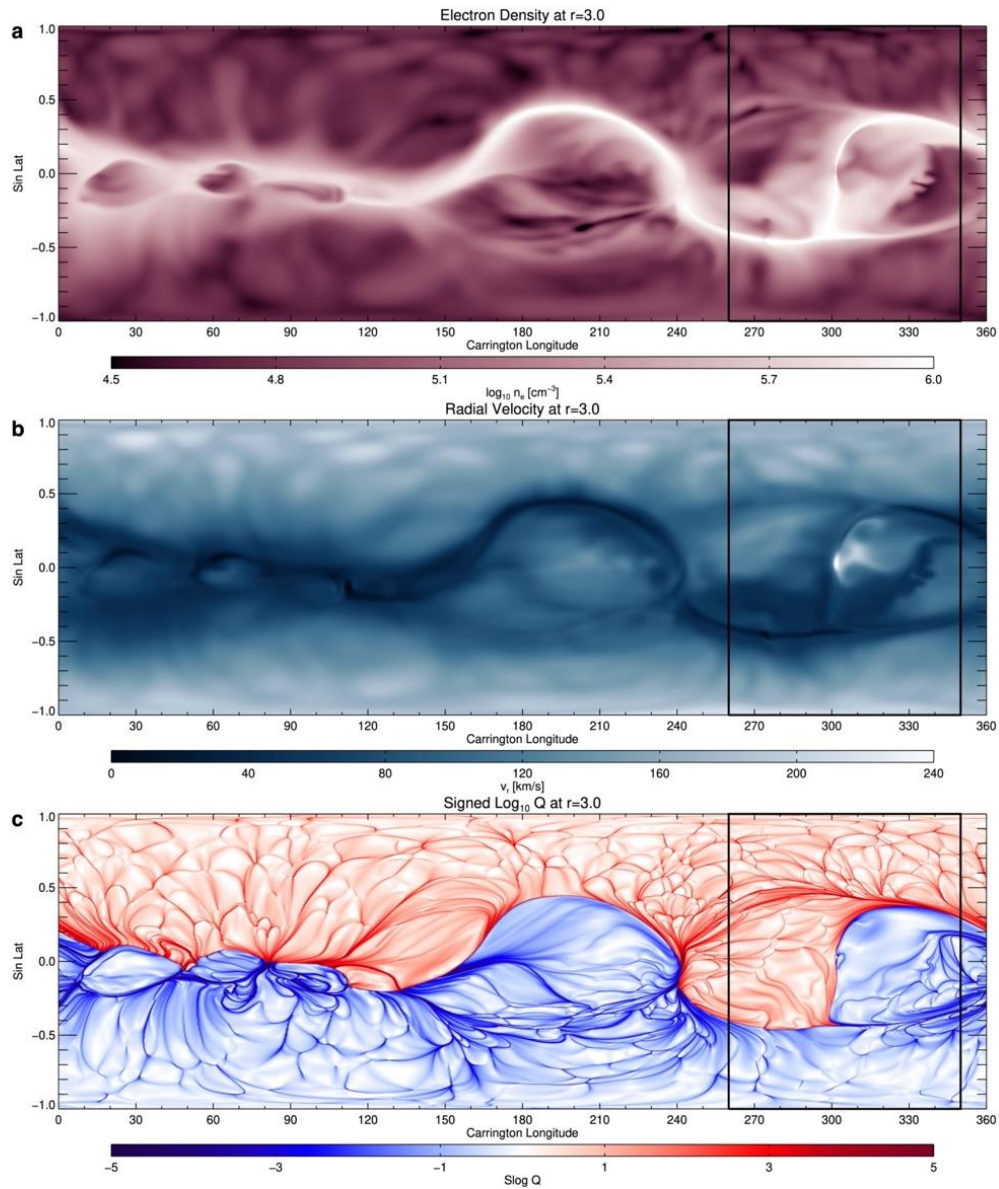

**Supplementary Figure 6: Plasma diagnostics and the S-web. a** and **b** show maps of electron number density and radial velocity from the MHD model sliced at $3R_\odot$. **c,** same as Extended Data Fig. 4c, showing synoptic signed $\log Q$ map at $3R_\odot$. The black box in all the panels covers the CH-AR system.

## Captions for Supplementary Videos

**Supplementary Video 1.** The larger panel in the bottom is same as the field of view displayed in Fig. 1c. Unlike Fig. 1c in which we show a sequence of running difference images, this larger panel in the video captures direct images. We zoom into a smaller sub field of view outlined (Fig. 1d) by a white rectangle to highlight the dynamics and interaction of structures in the complex coronal web. This video is associated with Fig. 1c-d.

**Supplementary Video 2.** Same format as Supplementary Video 1, but associated with Supplementary Figure 1c-d.

**Supplementary Video 3.** Same format as Supplementary Video 1, but associated with Supplementary Figure 2c-d.

**Supplementary Video 4.** Same format as Supplementary Video 1, but associated with Supplementary Figure 3c-d.

**Supplementary Video 5.** Still from right panel of this video are displayed in Fig. 1a and Figs. 2b-c. A still from the left panel of this video is displayed in Fig. 2a. The moving vertical line in the right panel marks the Carrington longitude of the west-limb. The right panel of the video captures the evolution of a complex coronal web (formed by low-latitude coronal hole active region system) and the emergence of highly structured slow solar wind streams over the west-limb for a period of roughly five days. The left panel shows the underlying magnetic topology at the west-limb, based on high-resolution 3D MHD simulations.

**Supplementary Video 6.** Format is similar to Supplementary Video 5. Stills from this video are displayed in Supplementary Figures 3a and 5. The video covers a period of roughly five days when the complex coronal web is over the east-limb.

**Supplementary Video 7.** A still from this video is displayed in Fig. 3a. The video showcases a 360° view of the 3D volume render of squashing factor. From Earth's perspective, the coronal hole active region complex appears closer to the disk center when the Carrington longitude of the central meridian (CML) is about 300°. The system appears on the west-limb for CML around 215°, and on the east-limb for CML around 25°.

**Supplementary Video 8.** A 360° view of the 3D volume render of squashing factor. FOV is ±6R☉ in X, ±3R☉ in Y. The CML indices that appear in the lower left of the image have the same meaning as in Supplementary Video 7. The video is associated with Fig. 3b. The solar disk is masked out in this video.

**Supplementary Video 9.** A still from this video is displayed in Fig. 4a. This STEREO composite image sequence showcases solar wind structures overlying the coronal hole active region system. The video covers a five-day period closer to the beginning of SUVI off-pointing observing campaign.

**Supplementary Video 10.** A still from this video is displayed in Fig. 4b. The format is same as Supplementary Video 9, but plotted for a period closer in time to the first perihelion of Parker Solar Probe (PSP) on 2018-11- 06. The stationary symbol points to an approximate plane-of-sky position of PSP at the time of its perihelion.